\documentclass{aa}  

\usepackage{graphicx}
\usepackage{txfonts}
\usepackage[pdfpagelabels=false]{hyperref}	
\hypersetup{colorlinks=true,linkcolor=blue,citecolor=blue,filecolor=blue,urlcolor=blue,}
\newcommand{\pcm}{~cm$^{-2}$}	
\newcommand{\pcmm}{~cm$^{-3}$}	
\newcommand{\kms}{~km\,s$^{-1}$}
\newcommand{\Hii}{H~{\sc ii}}

\usepackage{placeins}
\usepackage{multicol}
\usepackage{xcolor}
\newcommand{\iz}[1]{{\color{black} #1}}

\begin{document}

\title{A compact NH$_2$D source without NH$_3$ counterpart in the S187 high-mass star-forming region}


   \author{E. A. Trofimova\inst{1}
          \and
          I. I. Zinchenko\inst{1}
          \and
          C.~Henkel\inst{2} 
          \and 
          A. G. Pazukhin\inst{1}
          \and
          A. F.~Punanova\inst{3}
          \and 
          P.~M.~Zemlyanukha\inst{1}
          }

   \institute{Federal Research Center A.V. Gaponov-Grekhov Institute of Applied Physics of the Russian Academy of Sciences, \\ 46 Ul’yanov str., Nizhny Novgorod 603951, Russia\\
              \email{tea@ipfran.ru, zin@ipfran.ru
              }
            \and
Max-Planck-Institut f\"ur Radioastronomie, Auf dem H\"ugel 69, Bonn 53121, Germany 
              \and
            Department of Space, Earth and Environment, Chalmers University of Technology, Onsala Space Observatory, 43992, Onsala, Sweden 
             }

   \date{Received ...; accepted ...}

 
  \abstract
   {}
   {Our goal is to investigate a \iz{so far only} tentatively detected compact NH$_2$D source \iz{devoid of a notable} NH$_3$ counterpart, located rather close to the luminous IRAS 01202+6133 source (the projected distance is $\lesssim$0.3~pc) in the S187 high-mass star-forming region. A high ammonia deuteration in such an environment is unusual.}
   {We performed observations of this area with the 20-m OSO radio telescope in several frequency ranges from $\sim$72 to $\sim$112~GHz. The area was mapped at the lower frequencies while at the higher frequencies only the position of the NH$_2$D emission peak was observed. The data were analyzed using non-LTE modeling.}
   {The data confirm a compact NH$_2$D source with a size of $\lesssim$0.2~pc. In addition, emission from about 20 molecules was detected, but only the DCO$^+$(1--0) peak coincides with the NH$_2$D clump. The clump is not associated with any known astronomical source or an NH$_3$ emission peak but the dust continuum map shows an enhancement at approximately the same position. The gas kinetic temperature derived from various indicators is $\sim$20--30 K, which possibly characterizes the surrounding material. The intrinsic NH$_2$D line width ($\sim$0.4\kms) implies a clump with subsonic or transonic turbulence. The ortho/para NH$_2$D line intensity ratio is consistent with an abundance ratio of 3, expected from the spin statistics. The NH$_2$D/NH$_3$ abundance ratio may be $\sim$1. In general, the data hint at a compact cold gravitationally stable dense core within the warm environment, which may represent pristine material.
   However this picture still needs a verification with observations at a higher angular resolution.
   }
  {}


   \maketitle
%

\section{Introduction}
High mass star-forming (HMSF) regions are not only composed of hot freshly ionized gas forming H\,{\sc ii} regions and an active warm molecular environment, but there may also still remain  cool pockets of dense molecular gas well  enough shielded to exhibit chemical properties, which strongly differ from those of the environment. This component, potentially representing pre-collapse gas \citep[e.g.][]{Minh2018}, deserves particular attention.
\iz{The recent ASHES (ALMA Survey of 70 $\mu$m-dark High-mass clumps in Early Stages) and QUARKS (Querying Underlying mechanisms of massive star formation with ALMA-Resolved gas Kinematics and Structures) surveys \citep{Sanhueza2019,Yang2026} provide the first systematic, large-sample comparison of starless core properties across distinct evolutionary stages of high-mass star formation. The combined sample comprises 324 candidate starless cores: 197 from early-phase infrared-dark clouds (IRDCs; ASHES) and 127 from evolved-phase infrared-bright clouds (IRBCs; QUARKS). Both datasets utilize matched-resolution (approximately 0.5--0.6 arcseconds) ALMA Band 6 continuum and line data at 1.3 mm.
In summary, observations from ASHES and QUARKS 
favor competitive-like dynamical accretion scenarios for high-mass star formation.
}

Good tracers of cold gas are deuterated molecules \iz{\citep[e.g.,][]{Fontani2011, Gerner2015}}.
In 2017--2018, we performed a survey of a large sample of high-mass star-forming regions with the 20-m Onsala radio telescope in the lines of several deuterated molecules and their hydrogenated isotopologues \citep{Trofimova20, Trofimova2024}. During this survey we tentatively detected a compact peak of the ortho-NH$_2$D emission in the S187 HMSF region far away from the NH$_3$ peak but rather close to the luminous IR source IRAS 01202+6133 (the projected distance is $\la$0.3~pc). The peak is not associated with any known astronomical object. Later this peak was covered with the IRAM-30m radio telescope observations \citep{Pazukhin2023}. However, it was located at the very edge of the map. 
With the 100-m Effelsberg radio telescope, weak NH$_3$(1,1) emission was detected without a clear peak at this position. The intensity of the NH$_3$(1,1) line was lower than or comparable to the intensity of the ortho-NH$_2$D line, which implied a very high degree of ammonia deuteration. 

The S187 (also known as Sh2-187, KR120, LBN 126.7-00.80) 
HMSF region at a distance of $\sim$1~kpc is well studied from optical to radio wavelenghts \citep[a detailed description is presented in][]{Zemlyanukha2022}. It contains ionized gas, molecular clumps at different stages of evolution and a fragmented atomic shell around its \Hii\ region. 
\iz{\citet{Arvidsson2011} found it to be a likely site for triggered star formation.}
A couple of outflows were reported in the area discussed here \citep{Salas1998, Jiang2001}. IRAS 01202+6133 ($M\sim 21$~M$_\odot$) with a luminosity of $L\sim 5600$~L$_\odot$ is embedded in the envelope of the S187 \Hii\ region. Surroundings of such objects are usually rather warm. 
Chemical models predict a rapid drop of the ammonia deuteration with increasing temperature at $T\ga 20$~K \citep[e.g.,][]{Roueff07}, which is confirmed by observations \citep{Pazukhin2023}. Therefore, a high ammonia deuteration in this area is unexpected.

\iz{Deuteration of ammonia and other molecules in HMSF regions has been investigated in several works \citep[e.g.,][]{Pillai2007a, Busquet2010, Fontani2015, Wienen2021, Trofimova20, Trofimova2024, Pazukhin2023, Pazukhin2024}. 
\citet{Pillai2007a} derived NH$_2$D/NH$_3$ ratios in the range 0.004--0.67 for a sample of pre/protocluster clumps found in 32 HMSF regions.
Targeted case studies, such as G35.20-0.74, revealed extreme ammonia deuteration in its coldest subcores (kinetic temperatures $\sim 15$ K), with NH$_2$D/NH$_3$ ratios reaching levels of order unity after opacity corrections \citep{Busquet2010}.
\citet{Fontani2015} derived this ratio in the range from $\sim$0.01 to $\sim$1.4.
\citet{Wienen2021} found a large range of the NH$_2$D/NH$_3$ column density ratio up to 1.6$\pm$0.7 for a large sample of massive ATLASGAL (APEX Telescope Large Area Survey) clumps.
\citet{Pazukhin2023} obtained this ratio in the range 0.02--0.4 in five HMSF regions. \citet{Trofimova2024} derived the NH$_2$D/NH$_3$ ratios in the range $\sim$0.1--1 for a sample of 29 NH$_2$D sources detected in HMSF regions.
}

In 2023, follow-up observations with the 20-m Onsala radio telescope were carried out toward this region. The goal was to clarify the position of this object and determine its physical parameters. 
The remaining paper is organized as follows. In Sect.~\ref{sec:obs} we describe the observations, in Sect.~\ref{sec:res} we present and discuss the results, and draw our conclusions in Sect.~\ref{sec:concl}. 
Preliminary results have been presented in \citet{Trofimova2024maeu}.

\section{Observations and data reduction} \label{sec:obs}
The observations of the newly discovered  ortho-NH$_2$D peak were performed with the 20-m radio telescope of the Onsala Space Observatory (OSO) in April-May 2023 in the framework of the project O2022b-03. 
They were carried out with the 3~mm and 4~mm receivers \citep{OSO-3mm, OSO-4mm} in several frequency ranges: 71.8--74.2, 83.8--86.2, 97.1--99.5 and 109.1--111.5~GHz with the channel width of 76~kHz (equivalent to 0.21--0.27~\kms\ at these frequencies). In the two lower frequency bands, which cover the ortho-NH$_2$D transition at 85.9~GHz, an area of $3^{\prime} \times 3^{\prime}$ with the center position at R.A.(J2000) = 01$^\mathrm{h}$23$^\mathrm{m}$33{\fs}00, Dec.(J2000) = +61\degr47\arcmin50{\farcs}0 was mapped with the grid spacing of 20$^{\prime\prime}$. It was found that the emission peak of ortho-NH$_2$D is located approximately at the position (--20$^{\prime\prime}$, +20$^{\prime\prime}$).
In the two upper frequency bands \iz{covering the para-NH$_2$D transition at $\sim$110~GHz} only this position was observed with a long integration time. The observations were performed in the position switching mode with the reference position at R.A.(J2000) = 01$^\mathrm{h}$24$^\mathrm{m}$19{\fs}87, Dec.(J2000) = +61\degr42\arcmin17{\farcs}2, which was checked to be free of contaminating emission.
The half-power beam width (HPBW) was from $\sim$50\arcsec\ to $\sim$35\arcsec\ at the frequencies from 72 to 110~GHz, \iz{$\sim$43\arcsec\ at the frequency of the ortho-NH$_2$D transition}. The line intensities are converted to the main beam temperature scale ($T_\mathrm{mb}$) using main beam efficiencies ($\eta_\mathrm{mb}$) from 0.58 to 0.47 at the observed frequencies. The achieved rms noise in channels was $\sim$10~mK at the $T_\mathrm{mb}$ scale in the upper bands and $\sim$30~mK at the ortho-NH$_2$D frequency.

The data reduction was performed with the XS program developed by Per Bergman at OSO and with the GILDAS package\footnote{http://www.iram.fr/IRAMFR/GILDAS} developed at IRAM.
The non-LTE analysis of the data was performed with the RADEX program 
\citep{vdTak07}. 
The transition frequencies were taken from the 
CDMS \citep{CDMS2001, CDMS2005} and NIST\footnote{https://physics.nist.gov/cgi-bin/micro/table5/start.pl} databases. The CARTA program \citep{CARTA} was used to create the source maps. Kinetic temperatures were estimated from rotational diagrams using the CASSIS program \citep{CASSIS2015, CASSIS}.

\section{Results and discussion} \label{sec:res}

\subsection{Detected lines and maps}
\begin{figure*}
    \centering
    \includegraphics[width=0.52\linewidth]{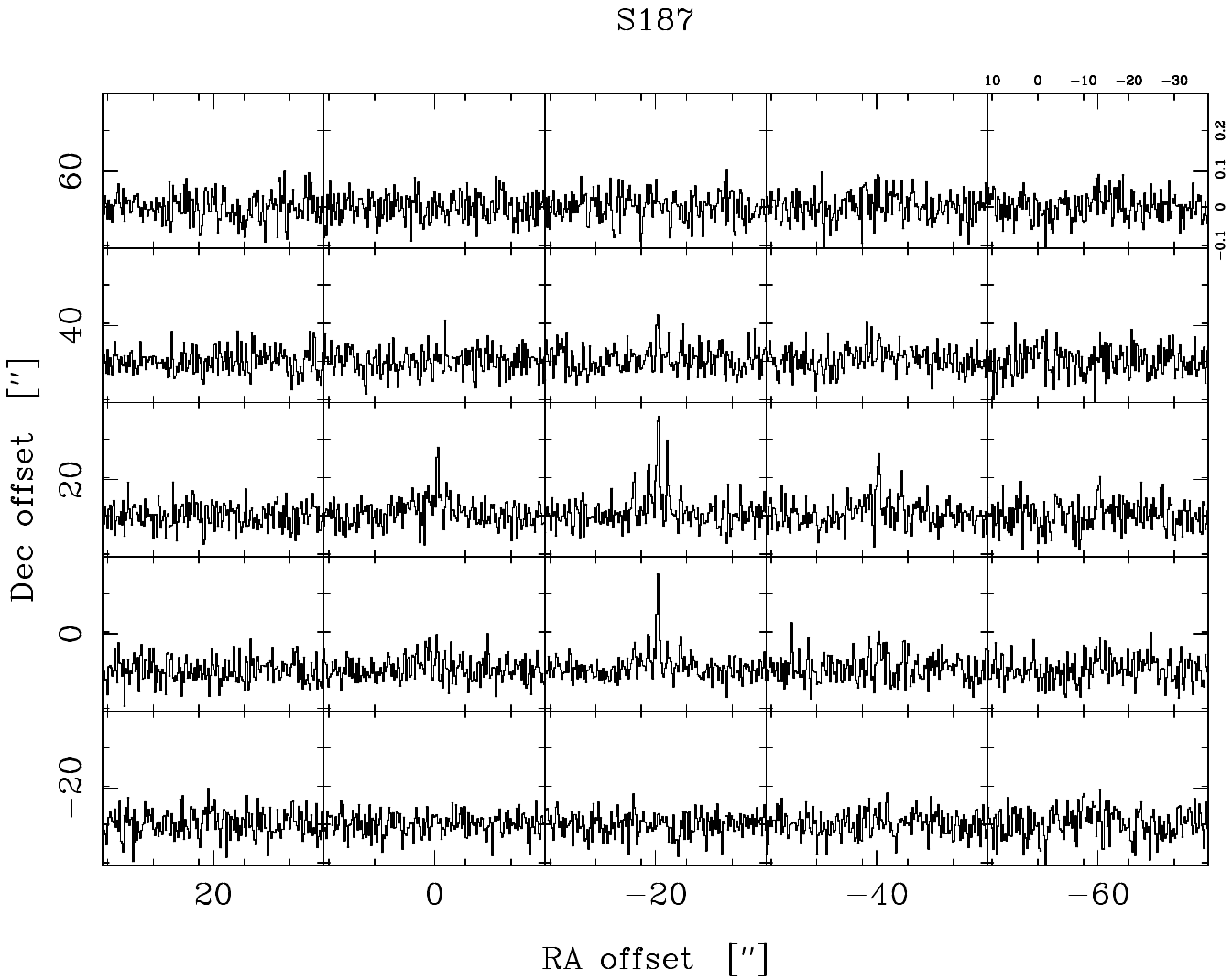}
    \hspace{1mm}
    \includegraphics[width=0.45\linewidth]{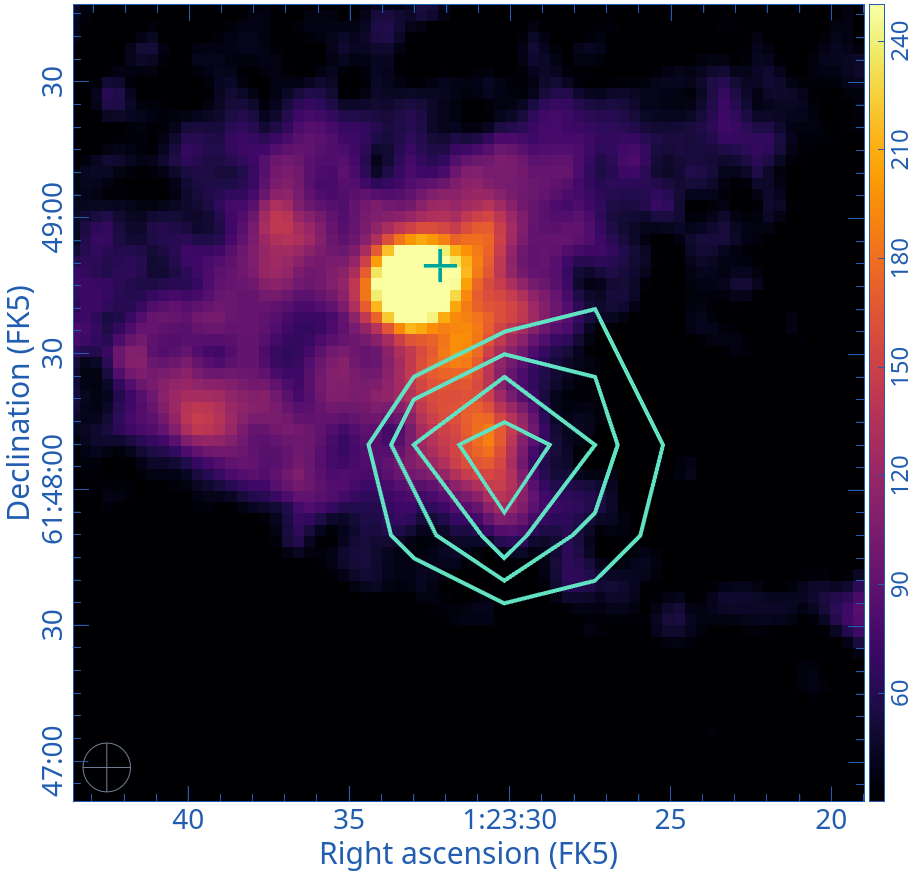}
    \caption{Left panel: raster map of the ortho-NH$_2$D emission toward the compact source discussed in this paper. The (0,0) position is at R.A.(J2000) = 01$^\mathrm{h}$23$^\mathrm{m}$33{\fs}00, Dec.(J2000) = +61\degr47\arcmin50{\farcs}0. Right panel: contours of the ortho-NH$_2$D emission overlaid over the 1.2 mm continuum image of the region \citep{Zin09}. The contour levels are from 0.08 to 0.2 by 0.04~K\,\kms. The intensity scale in the image is in mJy\,beam$^{-1}$. The cross marks the IRAS~01202+6133 source position. The beam of the 1.3 mm continuum observations is indicated in the lower left corner.}
    \label{fig:map}
\end{figure*}

In Figure~\ref{fig:map} we present the raster map of the ortho-NH$_2$D emission in the investigated area. A compact peak approximately at the position (--20$^{\prime\prime}$, +20$^{\prime\prime}$) is clearly seen in the left panel. In the right panel of this figure we overlay contours of the ortho-NH$_2$D emission over the 1.2 mm continuum image of the region obtained with the IRAM-30m telescope \citep{Zin09}.

The data confirm a compact source centered at R.A.(J2000) = 1$^\mathrm{h}$23$^\mathrm{m}$30{\fs}0, Dec.(J2000) = 61$^\circ$48$^\prime$06$^{\prime\prime}$ with a size comparable to or smaller than the beam size ($\sim$40$^{\prime\prime}$), which corresponds to $\la$0.2~pc at the distance to S187 of $\sim$1~kpc. The source is not associated with any known astronomical object but the dust continuum map shows an enhancement at approximately this position (see right panel in Fig.~\ref{fig:map}). The size of this feature is $\sim$20\arcsec, which corresponds to $\sim$0.1~pc.

Measurements of the NH$_3$ ($K,J$) = (1,1) line in this area were obtained in November 2019 with the 100-m telescope at Effelsberg (Germany)\footnote{The data were obtained with the 100-m telescope of the MPIfR (Max-Planck-Institut f{\"u}r Radioastronomie) at Effelsberg.} with a bandwidth of 300\,MHz and 65536 channels, providing a spectral resolution of $\sim$0.06\,km\,s$^{-1}$. These observations and their data reduction are described in \citet{Pazukhin2023}.
The beam width of the 100-m telescope at the NH$_3$(1,1) frequency ($\sim$40\arcsec) is rather close to the beam width of the OSO-20m antenna in the 3~mm band. The rms noise in channels was $\sim$0.2~K on the $T_\mathrm{mb}$ scale.

\begin{figure}
    \centering
    \includegraphics[width=\linewidth]{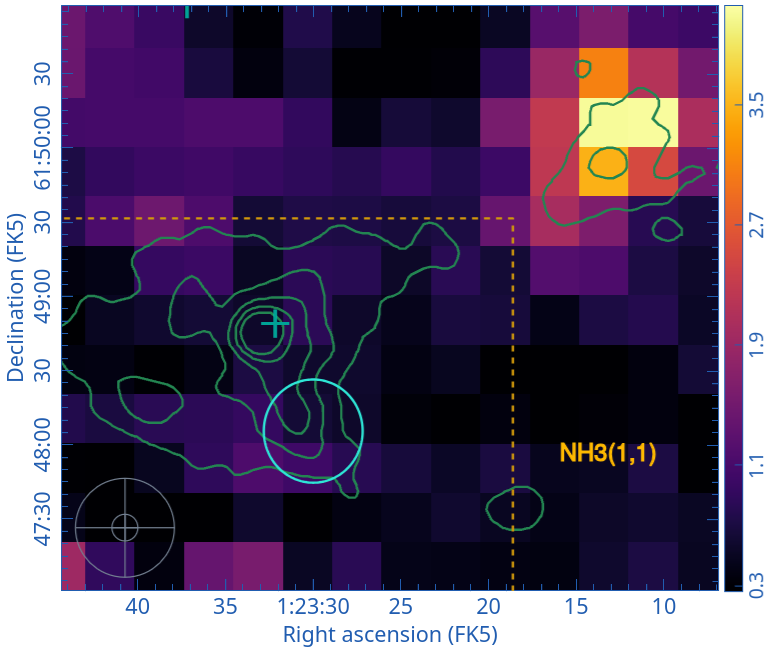}
    \caption{Map of the integrated intensity of \iz{all hfs components of} the NH$_3$(1,1) \iz{line} \citep[from the 100-m Effelsberg telescope, ][]{Pazukhin2023} with an intensity scale in units
    of K\kms\ (see the wedge on the right hand side). Overlaid are contours of the 1.2 mm 
    continuum emission \citep{Zin09}. The contour levels are from 50 to 250 in step of 50 mJy\,beam$^{-1}$. The cyan circle (40\arcsec in diameter) is centered at the position of the ortho-NH$_2$D emission peak discussed in this paper. The cross marks the same position as in Fig.~\ref{fig:map} (right panel). The beams of the NH$_3$(1,1) and 1.2 mm continuum observations are indicated in the lower left corner. The dashed line indicates boundaries of the region investigated in the present work.}
    \label{fig:nh3-map}
\end{figure}

In Figure~\ref{fig:nh3-map}, we present the map of the integrated intensity of all hfs components of the NH$_3$(1,1) line covering a somewhat larger area than in Fig.~\ref{fig:map}. One can see that there is no noticeable NH$_3$(1,1) peak toward the ortho-NH$_2$D emission peak. A slight enhancement of the NH$_3$(1,1) emission is observed $\sim$10\arcsec\ to the south of the ortho-NH$_2$D peak. However, similar features are seen in other parts of the map, too. They look as random variations of equally weak emission.

In addition, emission from about 20 molecules was detected 
(Table~\ref{table:lines}). In several lines, maps of the area \iz{could also be} obtained (Figure~\ref{fig:maps}). In these maps, only the DCO$^+$(1--0) peak coincides with the NH$_2$D clump. The HC$^{18}$O$^+$(1--0) line is too weak for constructing a map but it also shows a clear peak at the same position. Emission peaks of other molecules are either close to the IRAS source or are located about 1\arcmin\ farther to the east. An interesting exception is the c-HCCCH molecule. Its emission peak is located between the IRAS source and the NH$_2$D clump. Spectra of the detected lines are presented in Fig.~\ref{fig:all_spectra}.

\subsection{NH$_2$D and NH$_3$ line parameters}
\begin{figure}
    \centering
    \includegraphics[width=0.8\linewidth]{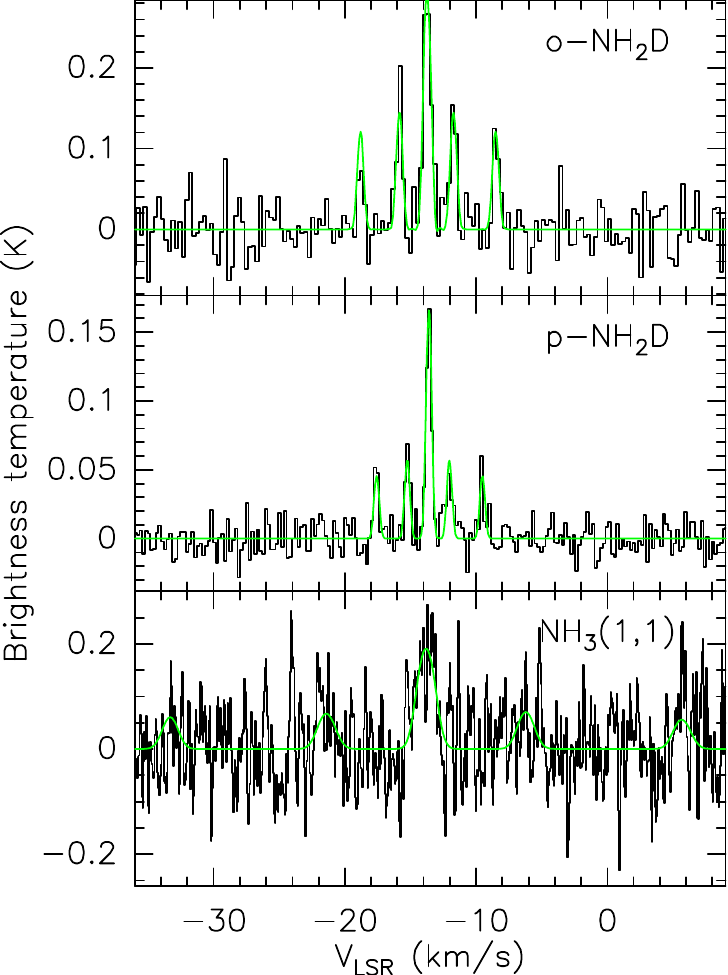}
    \caption{Spectra of the ortho-NH$_2$D, para-NH$_2$D and NH$_3$(1,1) (from top to bottom) toward the ortho-NH$_2$D peak. The NH$_3$(1,1) spectrum is averaged over the circle shown in Fig.~\ref{fig:nh3-map}. The GILDAS hfs fits are shown by the green curves.}
    \label{fig:spectra}
\end{figure}

In Figure~\ref{fig:spectra} we present spectra of the ortho- and para-NH$_2$D at the position (--20$^{\prime\prime}$, +20$^{\prime\prime}$) as well as the NH$_3$(1,1) spectrum averaged over the circle 40\arcsec\ in diameter centered at the ortho-NH$_2$D emission peak (Fig.~\ref{fig:nh3-map}) and their hfs fits .

The line widths ($\Delta V$) of individual hyperfine components determined by the HFS fitting in the GILDAS/CLASS program are 0.52$\pm$0.05\kms\ for ortho-NH$_2$D, 0.44$\pm$0.03\kms\ for para-NH$_2$D and 1.4$\pm$0.2\kms\ for NH$_3$(1,1). The optical depths of the main groups of the hfs components are 2.1$\pm$0.6 for ortho-NH$_2$D and 0.5$\pm$0.3 for para-NH$_2$D. 
The NH$_3$(1,1) spectrum is too noisy and an optical depth cannot be determined. The satellite components are hardly visible in the spectrum (lower panel of Fig.~\ref{fig:spectra}). The formal hfs fit suggests an optical depths of 0.9$\pm$0.7 for the main group of the hfs components.
\iz{The small difference in the line widths between ortho- and para-NH$_2$D is not significant with respect to the errors but might be slightly affected by different beam sizes at these frequencies, different critical densities for these transitions or by the higher opacity of the ortho line.}
Taking into account the spectral resolution, the intrinsic line width is $\sim$0.40\kms, which corresponds to a one-dimensional velocity dispersion of $\sigma \sim 0.17$\kms. 

\iz{This velocity dispersion is significantly larger than the thermal velocity dispersion for this molecule (0.07\kms\ at 10~K and 0.12\kms\ at 30~K). At the same time it is slightly lower than the isothermal sound speed in molecular gas  (0.19\kms\ at 10~K). \iz{We conclude} that the turbulence in this clump is subsonic or transonic.
}

\subsection{Physical properties and column densities in the NH$_2$D core and surroundings}
We estimated the gas kinetic temperature toward the NH$_2$D emission peak from the rotation diagrams for the symmetric top molecules CH$_3$CCH and CH$_3$CN. The result for CH$_3$CCH is $T_\mathrm{kin} = 31.7\pm0.8$~K. However, in Fig.~\ref{fig:maps} one can see that the bulk of the CH$_3$CCH emission originates not in the NH$_2$D clump but in the surroundings. In the CH$_3$CN $J_\mathrm{K}=6_\mathrm{K}-5_\mathrm{K}$ line series only the $K=0$ and 1 components are readily seen (Fig.~\ref{fig:all_spectra}). Using upper limits for higher components we obtain $T_\mathrm{kin}$\la20~K. The CH$_3$CN $J_\mathrm{K}=6_\mathrm{K}-5_\mathrm{K}$ emission was not mapped and its association with the NH$_2$D clump is unclear. The CH$_3$CN $J_\mathrm{K}=4_\mathrm{K}-3_\mathrm{K}$ line series at 73.6~GHz was not detected toward the NH$_2$D emission peak but it is rather strong in the area around the IRAS source and to the east of it. The gas kinetic temperature in the area of the NH$_2$D emission peak was estimated also from the HCN/HNC intensity ratio  as $T_\mathrm{kin} \sim$20--25~K \citep[Fig.~A2 in][]{Pazukhin2023}.

Additionally, we estimated the gas density toward the same position from non-LTE modeling of the two HC$_3$N and three SO lines in our bands (Table~\ref{table:lines}). The most reliable estimate can probably be obtained from the ($3_{2}-2_{1}$)/($2_{3}-1_{2}$) SO line intensity ratio since these lines are relatively close in frequency which diminishes possible calibration issues. This ratio gives density $n\sim 2\times 10^5$\pcmm\ with only a weak dependence on temperature. The ($2_{3}-1_{2}$)/($2_{2}-1_{1}$) SO line ratio gives a higher density $n\sim 5\times 10^5$\pcmm. From the integrated intensity ratio of two HC$_3$N transitions we obtain $n\sim 0.6\times 10^5$\pcmm\ at $T_\mathrm{kin} = 30$~K. However the line widths of these transitions are very different, which can be caused by the presence of a lower excitation component seen in several lines (Fig.~\ref{fig:all_spectra}). \iz{The ratio of their peak intensities yields} $n\sim 2\times 10^5$\pcmm. 
The derived column densities of SO and HC$_3$N are $N(\mathrm{SO}) \sim 2\times 10^{13}$\pcm\ and $N(\mathrm{HC_3N}) \sim 2\times 10^{12}$\pcm\ at $T_\mathrm{kin} = 30$~K.

We estimate the hydrogen column density from observations  of the C$^{18}$O line and from the dust emission. The LTE estimate of the C$^{18}$O column density (C$^{18}$O should be thermalized at the inferred density) at $T_\mathrm{kin} = 30$~K is $N(\mathrm{C^{18}O}) \sim 10^{16}$\pcm. Assuming a C$^{18}$O relative abundance of $1.7\times 10^{-7}$ \citep{Frer82} we obtain $N(\mathrm{H_2}) \sim 6\times 10^{22}$\pcm. 
Practically the same estimate of the hydrogen column density is obtained from the observed brightness of the dust emission assuming a dust specific absorption coefficient (per dust muss) of 0.9~cm$^2$\,g$^{-1}$ according to \citet{Ossenkopf94}, dust temperature of 20~K and a gas to dust mass ratio of 100. For a source size of 0.1~pc it implies a mean hydrogen density of $\bar{n}\sim 2\times 10^5$\pcmm. It practically coincides with the density in the regions of line formation found above from the excitation analysis.

The critical densities for the observed transitions are $n_\mathrm{cr} \approx 6.5\times 10^4$\pcmm\ for ortho-NH$_2$D and $n_\mathrm{cr} \approx 1.4\times 10^5$\pcmm\ for para-NH$_2$D at a temperature of 20~K as follows from the LAMDA database \citep{LAMDA}. 
Our modeling with RADEX \iz{using a uniform sphere model} shows that the ortho-NH$_2$D column density needed to match the derived optical depth in the ortho-NH$_2$D line is $N(\mathrm{o-NH_2D}) \sim \iz{3}\times 10^{14}$\pcm\ with a weak dependence on the adopted gas kinetic temperature and density in the ranges $T_\mathrm{kin} \sim$10--30~K and $n = (1-3)\times 10^5$\pcmm. \iz{In the RADEX model we fit the total optical depth of the transition which is twice the optical depth in the main group of the hfs components (since RADEX does not take into account the hfs splitting in NH$_2$D).} However, the ortho-NH$_2$D excitation temperature \iz{($T_\mathrm{ex}$)} under these conditions approaches the gas kinetic temperature and the beam filling factor should be low. For example, at $n = 2\times 10^5$\pcmm\ it should be $\sim$0.06 for $T_\mathrm{kin} = 10$~K \iz{($T_\mathrm{ex}\approx 8.8$~K)} and $\sim$0.02 for $T_\mathrm{kin} = 30$~K \iz{($T_\mathrm{ex}\approx 19$~K)} which implies the source size $\sim$6\arcsec--10\arcsec.

An estimation of the para-NH$_2$D column density from the derived optical depth would be hampered by a high uncertainty in this optical depth. Instead, we can estimate it from the observed line intensity assuming the same source size as for ortho-NH$_2$D. For a source much smaller than the beam at both frequencies, the beam filling factor for para-NH$_2$D will be higher than for ortho-NH$_2$D by a factor of 1.6 \iz{as follows from the difference in the beam widths}. The ratio of the observed peak intensities of the ortho- and para-NH$_2$D lines is 1.6$\pm$0.3. For a point-like source the ratio of the brightness temperatures in these lines should be $\sim$2.6$\pm$0.5 \iz{taking into account the difference in the beam filling factor}. According to the spin statistics, the abundance ratio of ortho- and para-NH$_2$D equals to 3 \citep{Sipila15}. Our modeling with RADEX in the same ranges of gas kinetic temperature and density shows that the observed intensity ratio is consistent with this value, especially at $T_\mathrm{kin} \la$20~K and $n \la 2\times 10^5$\pcmm. The excitation temperatures for ortho- and para-NH$_2$D are \iz{noticeably} different. \iz{For para-NH$_2$D $T_\mathrm{ex}\approx 6.9$~K at $n = 2\times 10^5$\pcmm\ and $T_\mathrm{kin} = 10$~K (compare with $T_\mathrm{ex}\approx 8.8$~K for ortho-NH$_2$D under the same conditions). This difference significantly affects the ratio of line intensities.}

Therefore, we adopt the ortho-to-para abundance ratio of three for NH$_2$D. Then, the total NH$_2$D column density is $N(\mathrm{NH_2D}) \sim \iz{4}\times 10^{14}$\pcm\ and the relative NH$_2$D abundance in the clump is $X(\mathrm{NH_2D})\ga \iz{6}\times10^{-9}$. It is a lower limit because the C$^{18}$O and dust emissions originate not only in this clump. This is close to the highest values observed in such regions \citep{Pazukhin2024, Trofimova2024}.

\iz{Another way to understand the low intensities of the NH$_2$D lines} can be an assumption of a very low clump density, $n\la 10^4$\pcmm, which can provide a sufficiently low NH$_2$D excitation temperature. However, existence of such a low density gas within a much denser environment is not likely. In principle, it could be adjacent to the dense material, but it is not clear how to explain its compactness and how it could maintain its low pressure. So a cold dense clump within a warm environment  is most plausible.


We estimate the NH$_3$ column density in two ways. First of all, it should be noted that the optical depth in the observed NH$_3$(1,1) line is apparently low. A substantial optical depth would imply a much higher brightness temperature. The critical density for this transition is only $n_\mathrm{cr} \sim 2\times 10^3$\pcmm. An assumption of a low beam filling factor is not supported by the observed distribution of the NH$_3$(1,1) emission (Fig.~\ref{fig:nh3-map}). However, we cannot exclude the presence of a significantly brighter and perhaps optically thick compact NH$_3$(1,1) source at the NH$_2$D peak location with an integrated line intensity, say, \la1/3 of the observed one. This possibility is discussed later, too.

Assuming an optically thin NH$_3$(1,1) source, we estimate the NH$_3$ column density from the integrated intensity of the main group of hyperfine components using the approach described in \citet{Harju93}. In this way we obtain $N(\mathrm{NH_3}) \sim 2.7\times 10^{13}$\pcm\ at $T_\mathrm{kin} = 10$~K and $N(\mathrm{NH_3}) \sim 1.7\times 10^{13}$\pcm\ at $T_\mathrm{kin} = 30$~K.
Another approach is based on RADEX modeling. In this way we obtain the column density of para-NH$_3$ of $N(\mathrm{para-NH_3}) \sim \iz{1.2}\times 10^{13}$\pcm\ with a weak dependence on temperature. 
Then, the total NH$_3$ column density is $N(\mathrm{NH_3}) \sim \iz{2.4}\times 10^{13}$\pcm\ assuming an ortho-to-para abundance ratio of unity as expected in thermal equilibrium at temperatures \ga30~K \citep{Faure2013}. This ratio can be much higher ($\sim$10) if the gas temperature at the time of NH$_3$ formation is very low \citep{Nishitani2012}. 
In general, both approaches give similar results indicating an NH$_3$ column density an order of magnitude lower than the NH$_2$D column density derived above. 

However, as mentioned above, we cannot exclude the presence of a significantly brighter compact NH$_3$(1,1) source. The constraints for such a source can be expressed as $fT_\mathrm{B}$\la 0.2~K and $\Delta V \sim 0.5$\kms, where $f$ is the beam filling factor and $T_\mathrm{B}$ is the source brightness temperature. Earlier we derived the beam filling factor for the ortho-NH$_2$D emission. It is $\sim$0.03--0.06. Assuming the same value for NH$_3$(1,1), we obtain $T_\mathrm{B}$\la 3--7~K for the NH$_3$(1,1) compact source. This source can well be optically thick, which requires $N(\mathrm{NH_3}) \ga \iz{7\times} 10^{14}$\pcm. In this case the limit on the brightness temperature translates also to the limit on gas kinetic temperature: $T_\mathrm{kin}$ \la 6--10~K.


This analysis shows that the most plausible interpretation of the available data is a compact cold dense core embedded in a warmer environment. Let us consider its stability. This is usually based on estimates of the virial parameter \citep{Bertoldi1992}:
\begin{equation} \label{eq:virpar}
    \alpha \equiv \frac{5\sigma^2R}{GM} \, ,
\end{equation}
where $R$ is the clump radius, $G$ is the gravitational constant and $M$ is the clump mass. Assuming a uniform sphere, it can be rewritten in the following forms:
\begin{equation} \label{eq:virpar_mod}
    \alpha \equiv \frac{15}{8\pi \ln 2} \frac{\Delta V^2}{GL N(\mathrm{H_2}) m_\mathrm{H} \mu_\mathrm{H_2}} = \frac{15}{8\pi \ln 2} \frac{\Delta V^2}{GL^2 \bar{n} m_\mathrm{H} \mu_\mathrm{H_2}} \, ,
\end{equation}
where $L$ is the clump size ($L=2R$), $m_\mathrm{H}$ is the H-atom mass and $\mu_\mathrm{H_2}$ is the molecular weight per hydrogen molecule, $\mu_\mathrm{H_2} = 2.8$ \citep{Kauffmann08}. 

The critical value of the virial parameter for gravitational stability is $\alpha \approx 2$ \citep{Kauffmann2013}. It corresponds to the following relation:
\iz{$LN\approx 1.4\times 10^{30}$~cm$^{-3}$s$^2$ $\Delta V^2$ $\approx 2.2\times 10^{39}$~cm$^{-1}$ $\approx 7.2\times 10^{20}$~pc\,cm$^{-2}$.}
This relation can be also expressed via the mean gas density: 
\iz{$L \bar{n}^{1/2}\approx 1.2\times 10^{15}$~cm$^{-3/2}$s $\Delta V$ $\approx 5\times 10^{19}$~cm$^{-1/2}$ $\approx 16$~pc\,cm$^{-3/2}$.}
For the mean density of $\bar{n}\sim 2\times 10^5$\pcmm, it implies a clump size of $\sim$0.04~pc and a beam filling factor for the Onsala observations of $\sim$0.03. Taking into account various uncertainties, it is sufficiently close to the above mentioned value for $T_\mathrm{kin} \sim 10-20$~K. The mass of this clump is $\sim$0.4~M\sun\ and $N(\mathrm{H_2}) \sim 2\times 10^{22}$\pcm. Then, $X(\mathrm{NH_2D})\sim \iz{2\times} 10^{-8}$. Such a size is in a good agreement with the well-known Larson's relation between size and velocity dispersion \citep{Larson1981}.

Our data do not allow us to estimate the ammonia content in this hypothetical compact core. \citet{Harju93} found the median value of the relative ammonia abundance in a sample of ammonia cores $X(\mathrm{NH_3})\sim 3\times 10^{-8}$. \citet{Ruoskanen2011} derived $X(\mathrm{NH_3})\sim 2\times 10^{-8}$ in the prestellar core Ophiuchus D (L1696A). \citet{Crapsi2007} obtained $X(\mathrm{NH_3})\sim 10^{-8}$ in the center of the dense core of L1544 and about 3 times lower value at the distance of $\sim 2\times 10^4$~au. Our derived $X(\mathrm{NH_2D})$ is comparable to these values indicating an ammonia deuteration fraction of $\sim$1.

Such a cold clump should be shielded from the radiation of the surrounding warm material. For the derived hydrogen column density and dust properties as adopted above, it will be opaque at wavelengths $\lambda \la 60$~$\mu$m, where most of the radiation from warm ($T\ga 30$~K) dust falls. A probable anisotropic distribution of such a radiation can also reduce its influence.

However, if the clump size is $\sim$0.1~pc, as observed for the continuum feature, it should be contracting since $\alpha \ll 2$. In such a case its existence would be much more enigmatic. The free-fall time for the derived density is $\la$10$^5$~years. Reliable measurements of the clump size and its physical properties are mandatory for solving this puzzle.

There are DCO$^+$ (Fig.~\ref{fig:maps}) and highly likely DNC \citep{Pazukhin2023} lines associated with this clump. The integrated intensity of the DCO$^+$ line is an order of magnitude higher than that of HC$^{18}$O$^+$. This indicates a deuterium fraction of \ga10$^{-2}$ which is not unusual \citep{Pazukhin2023}.
At the same time there is no noticeable DCN or N$_2$D$^+$ emission while deuteration of HCN and N$_2$H$^+$ is expected to be higher than that of HCO$^+$  \citep[e.g. Fig. 11 in][]{Nielsen2026arXiv}.
A low DCN/DNC ratio can indicate a low temperature. But the absence of the N$_2$D$^+$ emission is rather puzzling. It is expected to be abundant at the suggested conditions in this clump \citep[e.g.,][]{Tobin2013}. At the same time, its detection rate in our survey of HMSF regions in the lines of deuterated molecules \citep{Trofimova20} was very low. 

\section{Conclusions} \label{sec:concl}

The available data can be interpreted in the framework of a more or less self-consistent picture of a compact ($\sim$0.04~pc) cold ($\sim$10--20~K) quiescent gravitationally stable dense ($\sim 2\times 10^5$\pcmm) clump embedded in warm ($\sim$30~K) environment. It may represent pristine material remaining from the epoch preceding the star formation process in this area. However, there remains an important issue still to be solved. The solution and verification of the model as a whole require high-resolution interferometric observations. 
They can be instructive for a better understanding the process of high-mass star formation.


\begin{acknowledgements}
\iz{We are grateful to the anonymous referee for helpful comments.}
This work was supported by the Russian Science Foundation grant No. 24-12-00153 (https://rscf.ru/en/project/24-12-00153/).
We are grateful to Stan Kurtz for helpful discussions.
The authors acknowledge support from Onsala Space Observatory (especially Henrik Olofsson) for the provisioning of its facilities/observational support. The Onsala Space Observatory national research infrastrcuture is funded through Swedish Research Council grant No 2017-00648.

\end{acknowledgements}

%
\bibliographystyle{aa} 
\bibliography{s187} 
%
\onecolumn
\begin{appendix} 
\section{Detected lines} \label{sec:chmaps}

\begin{table*}[h!]
\caption{List of the detected lines and their parameters at the position (--20$^{\prime\prime}$, +20$^{\prime\prime}$) which corresponds to the ortho-NH$_2$D emission peak.
}             
\label{table:lines}      
\centering          
\begin{tabular}{c c c c c c c l }     
\hline\hline       
Frequency & Molecule & Transition & $E_{u}$ & $V_\mathrm{LSR}$ & $T_\mathrm{mb}$ & $\Delta V$ & Molecule\\
(MHz) &(formula) & & (K) & (\kms) & (K) & (\kms) & (name)\\
\hline                    
72039.31 & DCO$^+$ & 1--0 & 3.5 & --13.62(06) & 0.49(02) & 1.30(06) & Formylium \\
72758.24 & SO$_2$ & $6_{0,6}-5_{1,5}$ & 19.2 & --13.58(20) & 0.06(02) & 1.48(35) & Sulfur Dioxide \\
72783.82 & HC$_3$N & 8--7 & 15.7 & --13.66(03) & 0.47(02) & 1.49(08) & Cyanoacetylene \\
72837.95 & H$_2$CO & $1_{0,1}-0_{0,0}$ & 3.5 & --14.18(06) & 0.55(03) & 2.92(02) & Formaldehyde \\
 &  &  &  & --13.49(02) & 0.53(06) & 1.10(07) &  \\
84521.17 & CH$_3$OH & $5_{-1}-4_{0}$ & 40.4 & --13.59(20) & 0.07(02) & 2.32(46) & Methanol \\
85162.22 & HC$^{18}$O$^+$ & 1--0 & 4.1 & --13.51(04) & 0.16(03) & 0.45(07) & Formylium \\
85338.89 & c-HCCCH & $2_{1,2}-1_{0,1}$ & 6.4 & --13.46(03) & 0.31(03) & 0.82(09) & Cyclopropenylidene \\
85347.89 & HCS$^+$ & 2--1 & 6.1 & --13.48(15) & 0.09(02) & 1.34(42) & Thioformylium \\
85457.27 & CH$_3$CCH & $5_{0}-4_{0}$ & 12.3 & --13.58(06) & 0.18(02) & 0.94(14) & Methyl Acetylene \\
85926.26 & ortho-NH$_2$D & $1_{1,1}^s-1_{0,1}^a$ & 20.7 & --13.70(02) & 0.28(02) & 0.52(05) & Ammonia \\
86093.95 & SO & $2_{2}-1_{1}$ & 19.3 & --13.67(04) & 0.27(02) & 1.20(11) & Sulfur Monoxide \\
97171.84 & C$^{33}$S & 2--1 & 7.0 & --13.08(03) & 0.08(01) & 1.24(08) & Carbon Monosulfide \\
97301.21 & OCS & 8--7 & 21.0 & --13.20(10) & 0.03(01) & 1.30(20) & Carbonyl Sulfide \\
97715.39 & $^{34}$SO & $3_{2}-2_{1}$ & 9.1 & --13.41(02) & 0.14(01) & 1.15(04) & Sulfur Monoxide \\
97980.95 & CS & 2--1 & 7.1 & --15.28(01) & 0.61(01) & 1.56(02) & Carbon Monosulfide \\
 &  &  &  & --13.45(01) & 2.22(01) & 1.77(01) &  \\
99299.87 & SO & $3_{2}-2_{1}$ & 9.2 & --15.58(01) & 0.35(01) & 1.60(03) & Sulfur Monoxide \\
 &  &  &  & --13.79(02) & 1.68(01) & 1.29(01) &  \\
109173.63 & HC$_3$N & 12--11 & 34.1 & --13.25(01) & 0.37(01) & 0.83(03) & Cyanoacetylene \\
109252.22 & SO & $2_{3}-1_{2}$ & 21.1 & --13.25(01) & 0.29(01) & 1.15(03) & Sulfur Monoxide \\
109782.17 & C$^{18}$O & 1--0 & 5.3 & --14.73(01) & 1.12(03) & 0.81(02) & Carbon Monoxide \\
 &  &  &  & --13.28(01) & 5.35(02) & 1.34(01) &  \\
109905.75 & HNCO & $5_{0,5}-4_{0,4}$ & 15.8 & --15.56(20) & 0.02(01) & 1.47(20) & Isocyanic Acid \\
 &  &  &  & --13.43(05) & 0.07(01) & 1.09(13) &  \\
110153.59 & para-NH$_2$D & $1_{1,1}^a-1_{0,1}^s$ & 21.3 & --13.60(01) & 0.17(02) & 0.44(03) & Ammonia \\
110201.35 & $^{13}$CO & 1--0 & 5.3 & --14.66(02) & 6.86(13) & 2.16(04) & Carbon Monoxide \\
 &  &  &  & --13.35(01) & 8.75(16) & 1.71(01) &  \\
110383.50 & CH$_3$CN & $6_{0}-5_{0}$ & 18.5 & --13.54(13) & 0.03(01) & 0.90(29) & Methyl Cyanide \\
\hline                  
\end{tabular}
\end{table*}

\begin{figure*}[htb]
    \centering
    \includegraphics[width=0.245\linewidth]{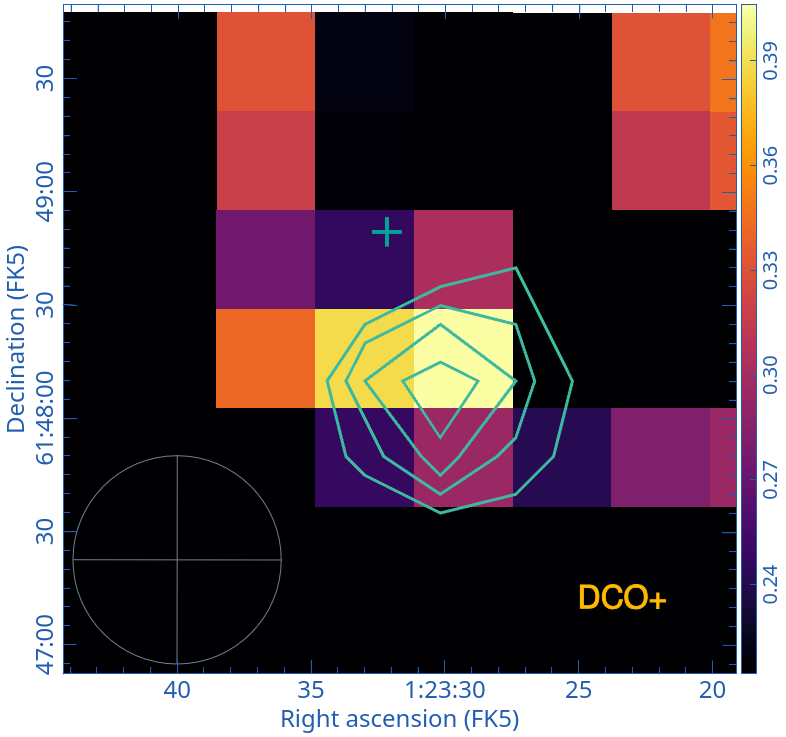}
    \hfill
    \includegraphics[width=0.245\linewidth]{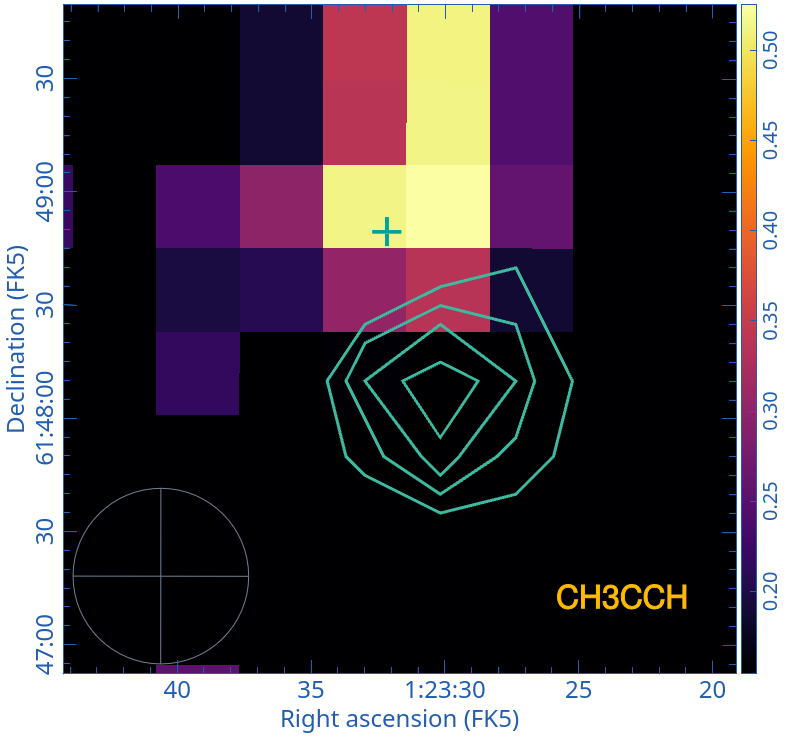}
    \hfill
    \includegraphics[width=0.245\linewidth]{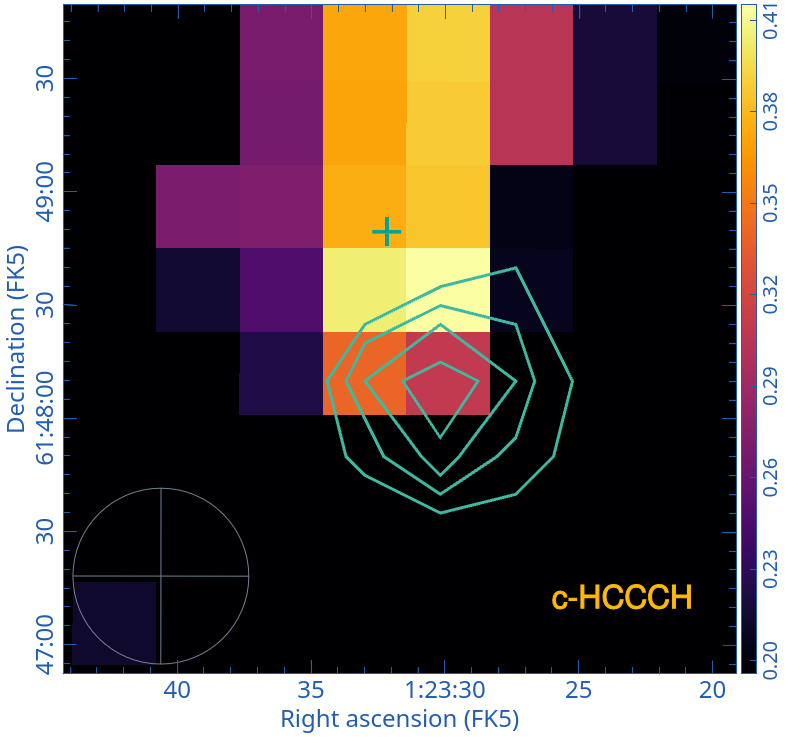}
    \hfill
    \includegraphics[width=0.245\linewidth]{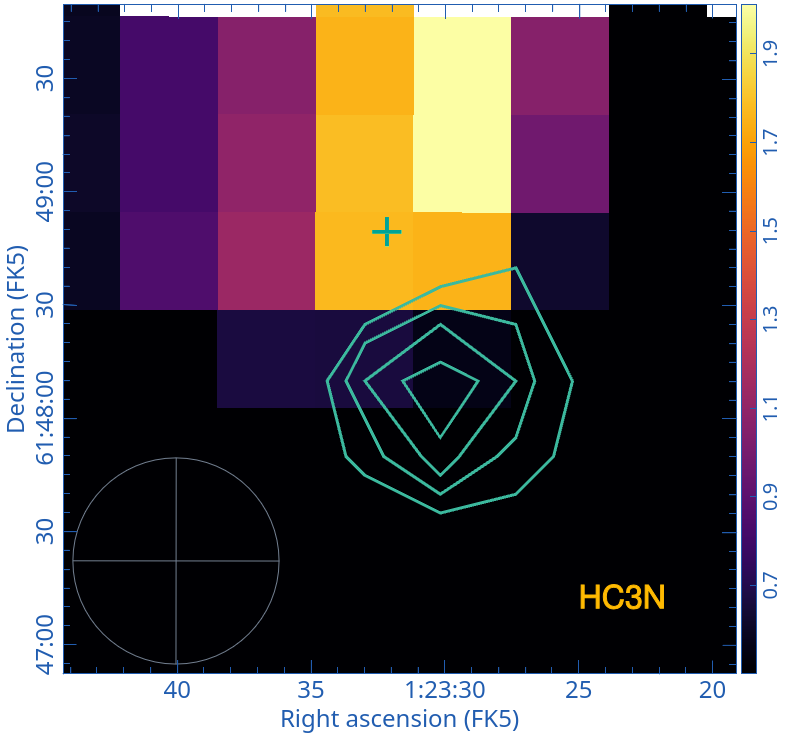}\\
    \includegraphics[width=0.245\linewidth]{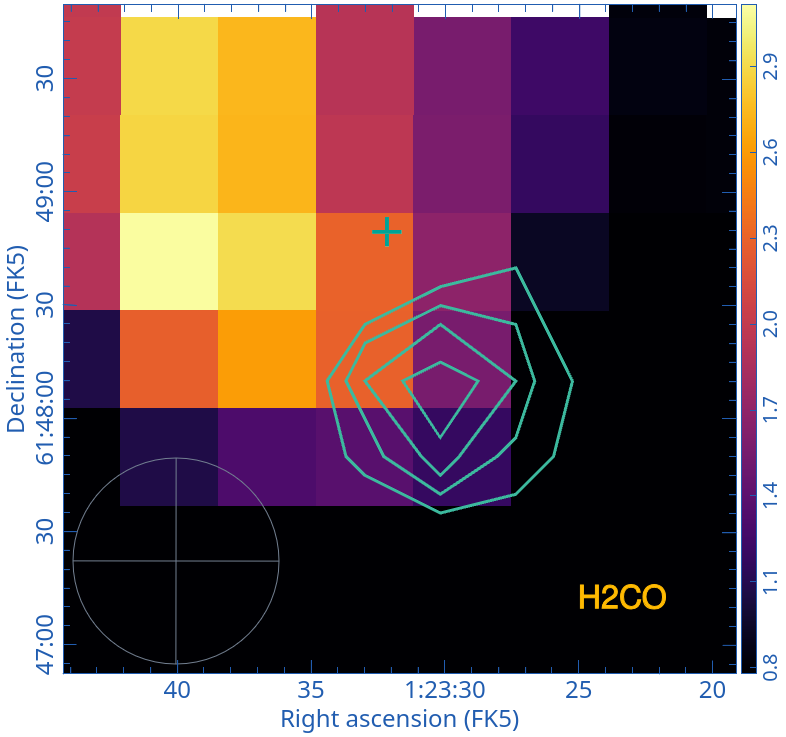}
    \hfill
    \includegraphics[width=0.245\linewidth]{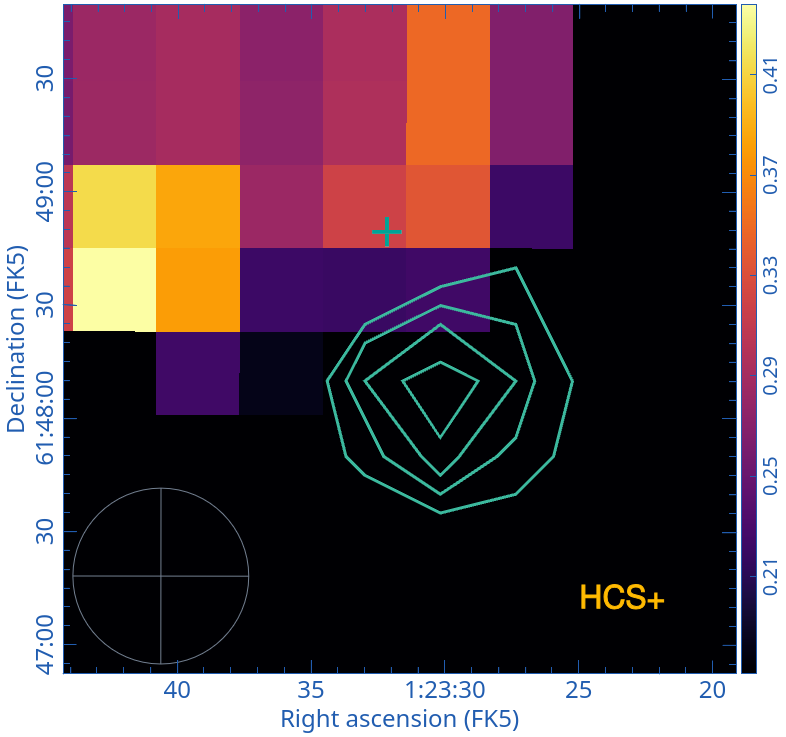}
    \hfill
    \includegraphics[width=0.245\linewidth]{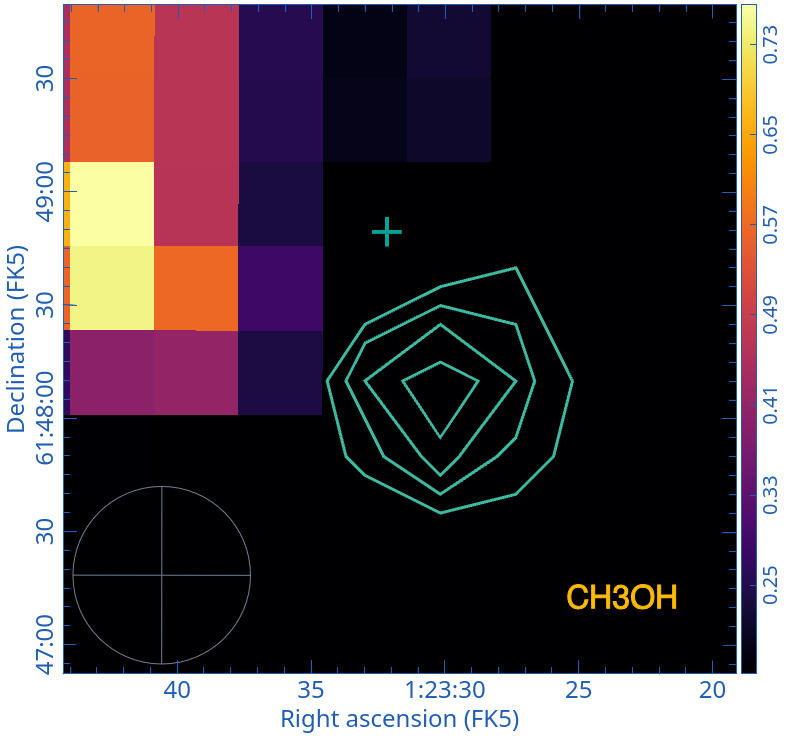}
    \hfill
    \includegraphics[width=0.245\linewidth]{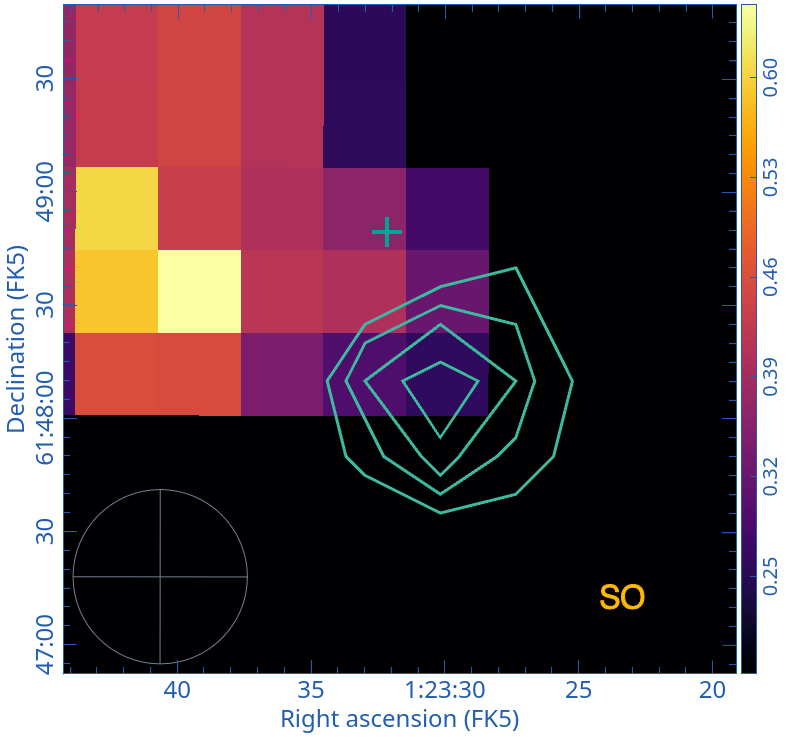}
    \caption{Maps of the integrated emission in several observed lines (indicated in the panels). The contours and the marker are the same as in Fig.~\ref{fig:map}.} 
    \label{fig:maps}
\end{figure*}

\begin{figure*}[htb]
    \centering
    \includegraphics[width=0.3\linewidth]{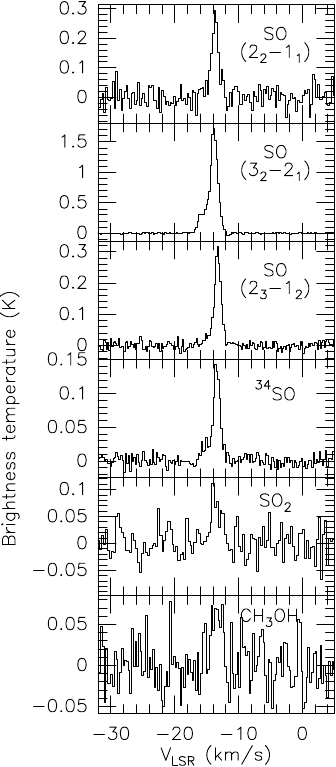}
    \hfill
    \includegraphics[width=0.3\linewidth]{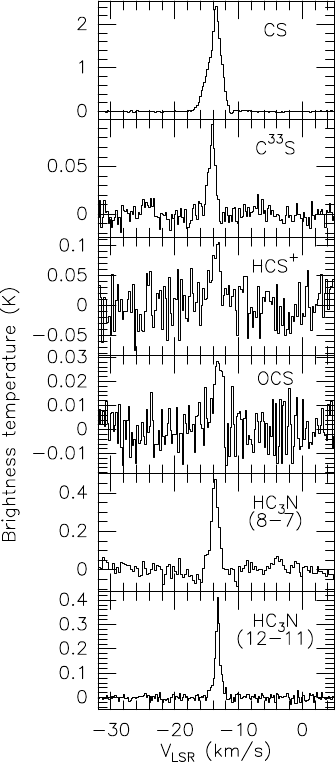}
    \hfill
    \includegraphics[width=0.3\linewidth]{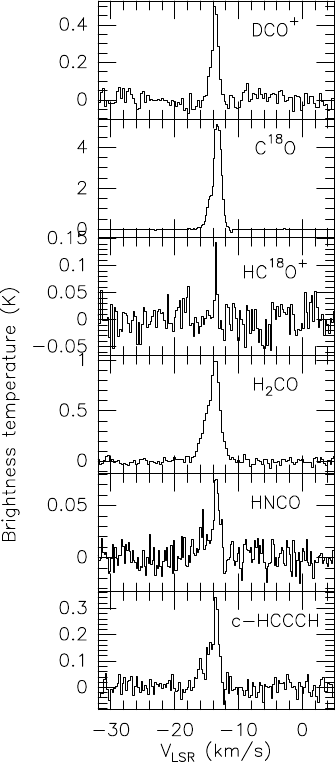} \\
    \vspace{3mm}
    \includegraphics[width=0.3\linewidth]{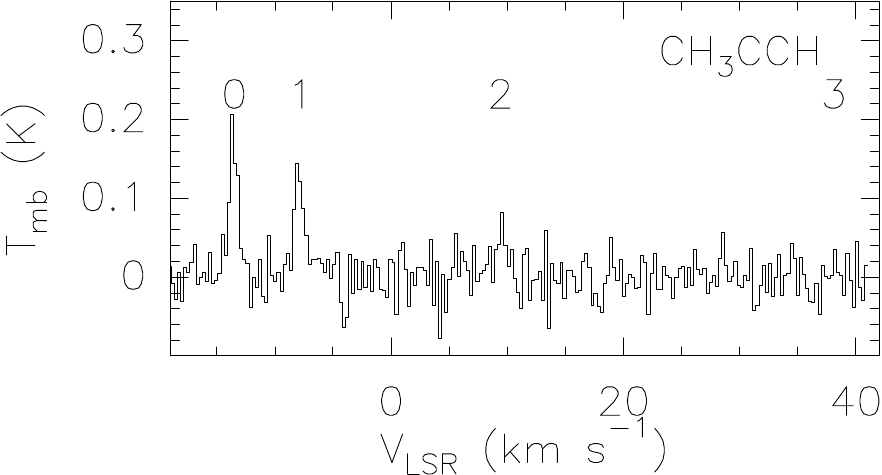}
    \hfill
    \includegraphics[width=0.3\linewidth]{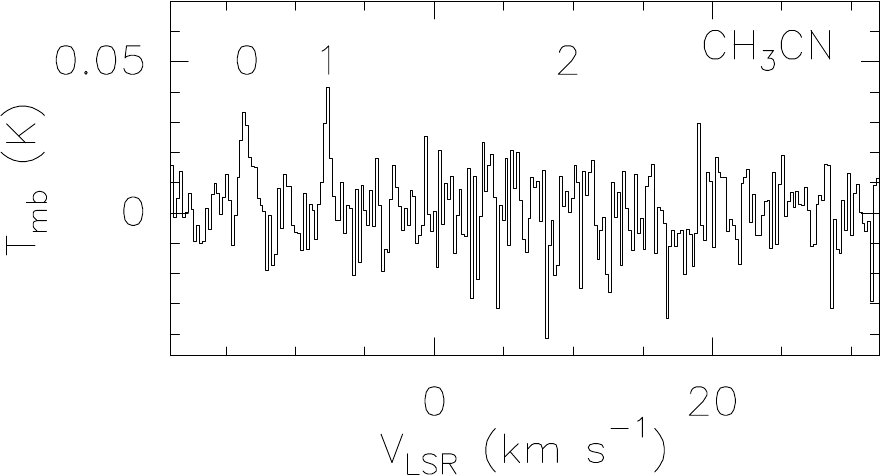}
     \hfill
    \includegraphics[width=0.3\linewidth]{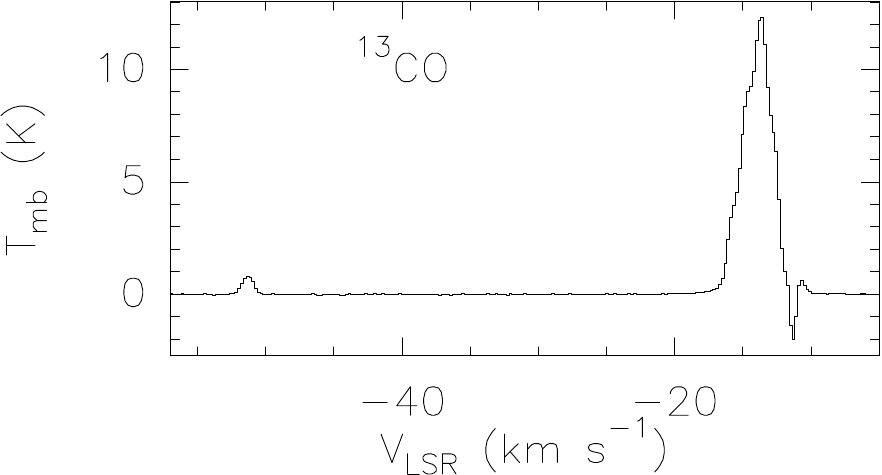} 
    \caption{Spectra of the detected lines at the position (--20$^{\prime\prime}$, +20$^{\prime\prime}$). For CH$_3$CCH and CH$_3$CN, locations of the $K$-components are indicated by corresponding numbers. The $^{13}$CO spectrum in the bottom also shows a component at $V_\mathrm{LSR} \sim -51$\kms. The emission of atomic hydrogen at such velocities was attributed to clouds in the Perseus arm \citep{Joncas1992} \iz{located well beyond the star forming region S187}.} 
    \label{fig:all_spectra}
\end{figure*}

\end{appendix}

\end{document}